\documentclass[11pt,onecolumn,letterpaper]{article}

\usepackage[spanish,english]{babel}
\usepackage[utf8x]{inputenc}
\usepackage[T1]{fontenc}
\usepackage[round]{natbib}
\usepackage{caption}
\usepackage{setspace}
\usepackage{comment}
\usepackage{fancyhdr}
\usepackage{tabularx}
\usepackage{xcolor}
\usepackage{amssymb}

\usepackage[top=2.6cm,bottom=2.6cm,left=2.6cm,right=2.6cm,marginparwidth=1.8cm]{geometry}
\usepackage{setspace}

\usepackage{amsmath}
\usepackage{graphicx}
\usepackage[colorinlistoftodos]{todonotes}
\usepackage[colorlinks=true, allcolors=blue]{hyperref}
\usepackage{float}
\usepackage{cleveref} 

\usepackage{sectsty}
\sectionfont{\fontsize{12}{15}\selectfont}
\subsectionfont{\fontsize{12}{15}\selectfont}
\usepackage[compact]{titlesec}

\fancypagestyle{specialfooter}{%
  \fancyhf{}
  
  \fancyfoot[C]{\fontsize{10}{0}\textit{Please cite this article as Acocella, Caplice, Sheffi, "Elephants or Goldfish?: An Empirical Analysis of Carrier Reciprocity in Dynamic Freight Markets", Transportation Research Part E: Logistics and Transportation Review, Vol 142, 2020, https://doi.org/10.1016/j.tre.2020.102073. \textcopyright 2020 This manuscript version is made available under the CC-BY-NC-ND 4.0 license https://creativecommons.org/licenses/by-nc-nd/4.0/}}
}

\fancypagestyle{number}{%
  \fancyhf{}
  
  \fancyfoot[C]{\thepage}
}

\begin{document}
{\centering{\Large Elephants or Goldfish?:\\ An Empirical Analysis of Carrier Reciprocity in Dynamic Freight Markets\par}\vspace{2ex}
{\large Angela Acocella, Chris Caplice, Yossi Sheffi
	\par}
	\smallbreak
     MIT\:Center\:for\:Transportation\:\&\:Logistics,\:1\:Amherst\:St.\:Cambridge,\:MA\:02142\par}\vspace{2ex}
        \noindent\rule{16.35cm}{0.4pt}
        \smallbreak

{\bfseries Abstract\par}
\smallbreak
Dynamic macroeconomic conditions and non-binding truckload freight contracts enable both shippers and carriers to behave opportunistically. We present an empirical analysis of carrier reciprocity in the US truckload transportation sector to demonstrate whether consistent performance and fair pricing by shippers when markets are in their favor result in maintained primary carrier tender acceptance when markets turn. The results suggest carriers have short memories: they do not remember shippers’ previous period pricing, tendering behavior, or performance when making freight acceptance decisions. However, carriers appear to be myopic and respond to shippers’ current market period behaviors, ostensibly without regard to shippers’ previous behaviors.
\medbreak
\textit{Keywords:} truckload freight transportation, tender acceptance, market cycles\\
\noindent\rule{16.35cm}{0.4pt}
\smallbreak

\thispagestyle{specialfooter}

\section{Introduction}
In the US freight transportation industry, shippers (e.g., manufacturers, wholesalers, distributors, or retailers with goods to be moved) and carriers (transportation service providers such as trucking companies that move those goods) interact constantly. These interactions between the two actors become important contributors to overall logistics costs. Over-the-road transportation moves approximately 10 billion tons of freight per year \citep{ATA2018} and accounted for 4.3\% of the \$19 trillion US GDP in 2017. The two for-hire trucking options available to shippers are full truckload (TL) - large shipments, typically over 10,000 pounds, where the shipper is usually the sole customer - and less-than-truckload (LTL) - smaller shipments that cannot fill an entire trailer where multiple shippers' loads are consolidated together. For-hire TL transportation spend totaled \$296.1 billion in 2018, or 44.3\% of over-the-road freight transportation \citep{CSCMP2018}.

As a result of deregulation of the industry following the 1980 Motor Carrier Act, shippers have followed two approaches to procure for-hire TL transportation. They can enter into long-term contracts with carriers or seek capacity on the spot market. To procure contracted capacity, shippers conduct a reverse auction\footnote{In a reverse auction, rather than the buyer responding to the seller with his willingness to pay for the good or service, the seller of the good or service responds to the buyer with the price she is willing to accept.}. They forecast expected annual transportation demand for each lane (origin-destination pair) within their freight distribution networks, decide if and how to package lanes together in order to make them more attractive to carriers, and invite carriers to participate by issuing a Request for Proposals (RFPs). Carriers respond to RFPs by stating the price they are willing to accept to serve each lane. This price is based on origin and destination characteristics (e.g. expected dwell time or likelihood of obtaining a follow-on load after drop-off), expected volume (number of loads offered per week), the carriers' operating costs, fleet size, and existing network. The shipper typically selects the lowest bidding carrier with which to enter a contract, subject to some level of service constraints; this is the primary carrier. Under these contracts, prices are set and held exclusively and privately between shippers and carriers, typically lasting one or two years. (See \cite{Caplice2007} for a detailed discussion of the TL transportation procurement process).

As an alternative to contracted services, shippers can have immediate demand served on the spot market - an open marketplace where prices are determined on a load-by-load basis and shippers and carriers are matched for individual load transactions. Typically, about 72\% of for-hire TL freight is accepted by primary carriers, while only 6\% is fulfilled by spot capacity \citep{Yuan2019}. As the industry evolved post deregulation, both shippers and carriers sought relationships with one another to encourage better performance and pricing consistency \citep{Crum1991, Gibson1993}.

These relationships are difficult to ensure, however, as the terms of the contract between shippers and primary carriers are non-binding in terms of volume and frequency commitments. This is a unique quality of TL transportation as compared to the procurement of other services. When shippers and carriers enter long-term contracts, for each origin and destination pair (lane), only the price at which the carrier agrees to provide services is binding; there is no hard and fast commitment on the shipper's part to offer loads to the carrier. Nor is there any commitment on the carrier's part to accept all loads when offered.

To address potential primary carrier load rejections, shippers compile a routing guide, consisting of an ordered list of backup carriers that had responded to the shipper's RFP. Typically these carriers' bid prices (i.e. price at which they indicated willingness to serve the lane) are higher than that of the chosen primary carrier. If a load is rejected by the primary carrier, the shipper's transportation manager consults the routing guide and offers the load to an alternate carrier, which again can accept or reject the load. In the case of the alternate carrier rejection, the shipper continues down the routing guide until one of the backup carriers accepts the load. If the routing guide fails, the shipper must turn to the spot market. Backup carriers' prices indicated during procurement are not binding as contracted carriers' are, and these backup carriers may demand higher prices if they are called upon later. \cite{Yuan2019} find that on average, backup carrier prices are 9-12\% higher than contract prices, depending on routing guide depth, while average spot rates exceed contract prices by 23-35\%, depending on market conditions.

The non-binding agreements further complicate planning long-term budgets and services accurately for both parties. The problems are compounded as shippers procure transportation services from many carriers across their distribution networks and carriers serve many shippers \citep{CapShef2003,CapShef2005}. Shippers must balance trade-offs between pricing, expected transportation service levels, and their own customer demand uncertainty, while carriers face a dynamic pickup and delivery problem in which they face random demand and short lead times as their trucks move from point to point \citep{Powell1978,Berb2010}\footnote{See \citet{Williamson1993,Williamson1985} for a detailed discussion of the trade-offs and implications of contract economics and relational contracts.}.

These issues worsen when macroeconomic factors are considered. The freight transportation industry undergoes market fluctuations due to changes in the overall economy, hedging or otherwise opportunistic pricing during the transportation procurement process, truck driver shortages or churn, regulations, and extreme weather events \citep{Coyote2018}. In what is referred to as a soft market, the demand for transportation services is less than available trucking capacity, driving prices down and carriers to accept only marginally profitable or even unprofitable freight – this is when shippers hold the more advantageous position. On the other hand, in a tight market, demand for transportation outstrips supply, carriers have their pick of freight they want to move, prices are high, and carriers have the advantage. The most recent cycle from soft to tight market occurred around Q3 2017 \citep{Coyote2018}, and markets began to soften again around the Q4 2018.

A key performance indicator used by shippers is a primary carrier's acceptance ratio (PAR) - that is, the fraction of loads the carrier accepts relative to the number of loads that it is offered by the shipper on a lane a carrier has won as primary carrier. When a primary carrier's PAR declines, as is seen in tight periods, shippers must go deeper in their routing guides to find non-contracted carriers to move their loads. Elevated freight rejection and high transportation costs in tight market periods hinders overall business performance \citep{Smith2019}. As a result, during tight periods shippers search for ways to develop relationships with their carriers and ensure consistent performance through the cycle, seeking `shipper of choice' status, \citep{CHR2015}.

The tensions between shippers and carriers that we highlight motivate this study, in which we test the often-heard opinion by shippers and carriers that they each ``have long memories'' between market cycles. 

\section{Literature Review}
A vast set of literature related to the PAR problem models the load matching and carrier assignment decisions to help shippers and carriers improve performance. Due to advances in near-real-time data availability and computational capabilities, much of the recent literature has explored dynamic versions of these problems, where the shippers' load information becomes available over time rather than being known a priori. A related, but much smaller body of literature studies shipper-carrier relationships.




\subsection{Strategic carrier assignment decisions}
Two problems in the transportation literature, which are related but distinct, are the shipper's strategic decision during the procurement process of which primary carrier to assign to each lane, and the operational decision regarding which carrier to offer each load.

\cite{Caplice1996} and \cite{CapShef2003} argue that economies of scope come about when shippers and carriers make procurement and bid decisions by considering their network as a whole rather than on a lane-by-lane basis. A combinatorial optimization approach to the procurement auction is developed to improve efficiencies; in particular, to improve how carriers communicate their cost function information to shippers and how shippers assign carriers to lanes. These studies focus on the bids that carriers submit in response to the RFP and develop formal optimization models and solutions for how shippers should analyze these submissions and assign carriers to lanes.

Building on this research, \cite{Lu2003} considers both bid cost and carrier reputation in the shipper’s combinatorial carrier assignment problem. The author finds that when taking reputation into account during the primary carrier selection process, shippers pay more in total direct costs to carriers, but there is a decrease in total hidden costs, which results in a net savings in total transportation costs. Including carrier reputation in the assignment decision is also considered by \cite{Rekik2012}. Here, the reputation-based winner determination problem where allocation of long-term contracts to carriers is decided based on both bid ask prices and carriers' reputation. The carriers' reputation is translated into unexpected hidden costs the shipper may incur when working with the winning carriers.

The above literature assumes the shipper has decided to outsource the transportation service and models the shipper's selection of for-hire carriers. However, a number of studies have approached the carrier assignment problem in which the shipper chooses between contracting for-hire carriers and serving the lanes itself with a private fleet. This set of literature includes \cite{Guastaroba2009}, \cite{Walmart2011}, \cite{Lee2007}, and \cite{Mulqueen2006}. For a review of the TL transportation service procurement modeling approaches including contract or spot agreements between shippers and carriers, the reader is referred to \cite{Basu2015}.

\subsection{Operational load matching decisions}
The load assignment problem finds the minimum-cost assignment of loads to carriers, where each load must be picked up at a given location at a specific time and delivered to a known destination (see \cite{Powell1995}). \cite{Powell1996} presents a methodology to evaluate the dynamic load assignment model with demand uncertainty and real time information using rolling horizon simulations.

Similarly, simulation approaches have been applied to study the dynamic nature of TL load matching operations by \cite{Tjokroamidjojo2006} and \cite{Regan1998}. Further, \cite{Simao2009} demonstrate an approximate dynamic programming method to accurately simulate the operations of the largest truckload motor carrier in the US, with over 6,000 drivers. The authors make use of both pre- and post-decision states in order to deal with the complexities of carriers’ operations. The model includes driver attributes such as home domicile, desire to be offered loads that get them back home, and hours already worked. Load attributes including origin and destination, appointment and delivery types, and revenue, are also included to allow the company to test the value of changes in the mix of drivers, freight, and other operating policies. The authors highlight that their model can be extended to help the carrier make tender acceptance decisions at the load level based on the current state of the system and ultimately evaluate how to commit to contractual agreements with shippers.

\subsection{Shipper-carrier relationships}
The S-C relationship as it relates to market conditions and the nature of the resulting contractual agreements is the focus of \cite{Hubbard2001a}, which studies relationship-specific investments in the trucking industry. The author argues that in trucking, the customer (i.e., shipper) rarely experiences asset specificity - that is, investments specific to partners (carriers) over long horizons as compared to other industries. He argues that with increasing long-haul trucking market thickness - that is, as more buyers and sellers come to the market - shippers and carriers utilize simple spot transactions rather than more complicated contractual arrangements.

Much of the extant transportation literature on shipper-carrier (S-C) relationships relies on surveys, interviews, and isolated experiments rather than empirical industry data, in part due to the difficulty of obtaining private company data. Carriers' freight acceptance has been the focus of a few studies based on transaction data, however. Rather than measures of the S-C relationship, most authors consider attributes of the lanes and freight in determining whether a load will be accepted. High lane volume \citep{Harding2005}, low lane volume volatility \citep{Kim2013}, high pricing \citep{Amiryan2015}, and high lane consistency, or cadence, \citep{Yuan2019} have been found to be positively correlated with higher primary carrier AR (PAR).

The S-C relationship, however, has been found to contribute to carriers' freight acceptance in a few studies. \cite{Zsidisin2007} find that contracted carriers, which shippers believe represent good relationships, outperform carriers that demonstrate poor or purely transactional relationships in terms of freight acceptance, on-time delivery, and pre-positioned capacity. 

\cite{Scott2017} analyze contract and spot market transactions and the impact the S-C relationship on freight acceptance. They find that less frequent load offers increase the likelihood of a carrier rejecting a load, while higher offered volume, lower load offer volatility, and higher revenue transacted between the shipper and carrier increase the likelihood of carrier's load acceptance. The authors do use a measure of market condition - Spot Premium - which is the ratio of 7-day average spot prices to 7-day average contract prices in a geographic region. As the unit of analysis in the model is at the load level, this rolling ratio still indicates immediate market conditions, testing whether carriers are more or less likely to accept loads in one market or another. We extend this market condition consideration by including \textit{previous} market behaviors into the acceptance decision model.

Each of these studies utilizes data from a single shipper, which limits their ability to generalize across types of shippers or to segment the data by S-C relationship types. Moreover, while the above literature explores the S-C relationship and its effects on carriers' tender acceptance, it does not consider previous behaviors explicitly defined by market conditions. Our research contributes to the TL transportation literature in that we analyze transactions and behaviors between many US shippers and contracted carriers and determine how the relationship between them in one market condition corresponds to performance in the next, when power dynamics have shifted.

\section{Research Question and Hypotheses}
To address the research gaps described above, we formulate the following research question, which considers the degree to which shippers and primary carriers stand by the terms of their previously defined non-binding contracts as market conditions (and power dynamics) change:\\
\\
\textbf{RQ: Should shippers pay competitive prices or provide high, consistent freight volumes when markets are soft to ensure carriers will adhere to contract expectations when the market tightens with high acceptance ratios of contracted loads?}\\

In particular, do carriers have long memories (thus, act like elephants) and reciprocate high soft market period pricing, tender consistency, or high volume from shippers with consistent, high PAR in tight periods? Conversely, do carriers reciprocate low soft period pricing, tender consistency, or volume from shippers with lower tight period PAR? Or, do carriers act like goldfish with short memories, forgetting shippers' previous behaviors?

In short, our research question considers whether it is worth it for a shipper to ``pay it forward'' by strictly adhering to contract commitments even when lower cost options may be available. This we measure with primary carrier acceptance ratio and formulate the following hypotheses:
\subsection{Carriers' response to \textbf{previous} shipper behavior}
First, we focus our attention on whether carriers respond to shippers' previous behaviors. That is, do carriers have long memories (thus, act like elephants) and reciprocate good performance soft market from shippers with high PAR in tight periods? We measure ``good'' performance as competitive pricing, consistent tendering patterns, advanced load notice, minimal delays at pick-up and delivery, and frequent business. We consider asset-based primary carriers independently from non-asset primary carriers (which, in our context includes brokers serving as contracted carriers), as they may have different strategies, access to capacity, and perhaps tolerances for certain customer behaviors. This leads us to the following two hypotheses:

\textbf{H1a: } Asset-based primary carriers offer high tight market period PAR for their shippers that had demonstrated good performance in the previous soft market period.

\textbf{H1b: } Non-asset primary carriers offer high tight market period PAR for their shippers that had demonstrated good performance in the previous soft market period.

\subsection{Carriers' response to \textbf{current} shipper behavior}
Second, we consider whether carriers respond to shippers' current behaviors. That is, do carriers act more like goldfish and offer high tight market PAR for shippers demonstrating current good behaviors? Thus, we formulate the following two hypotheses:

\textbf{H2a: } Asset-based primary carriers offer high tight market period PAR for their shippers that demonstrate good performance in the current tight market period.

\textbf{H2b: } Non-asset primary carriers offer high tight market period PAR for their shippers that demonstrate good performance in the current tight market period.\\

\section{Methodology}
In the following section, we describe the hybrid break point detection method implemented for this study, define the market time periods, characterize the S-C pairs and describe the model we implement to answer our research question.

\subsection{Empirical Data Summary}
The data for this research is obtained from a partner company that provides logistics and transportation functions for its shipper clients. Our partner company is not a brokerage, but rather is a managed services provider assisting its clients (shippers). The relationship and all payments are directly between the shipper and the carrier with no intermediaries. This is important to note, as we intend to study the direct relationship between a shipper and its primary carriers, not how carriers interact with a third party. Many companies offer such managed transportation services. For example, Transplace, BluJay Solutions, 4Flow, C.H. Robinson, and XPO Logistics. While the latter firms are two of the largest traditional 3PL providers in the US, they both offer independent managed transportation services as described above without promoting or relying on the brokerage side of their businesses.

For our purposes, we obtain both tactical pricing information and operational load transaction data. The carriers' pricing information is obtained during each shippers' procurement event; prices carriers bid for each lane are collected, one or more carriers are selected to be contracted, and a subset of the remaining carriers that bid are retained as backup carriers in cases where the primary carriers reject loads. Typically this process is either done solely by the client shipper, or with the help of our partner company, depending on the the shipper's needs. This pricing information for each carrier - primary (i.e., contracted) and all backup (i.e., non-contracted) - is uploaded into the routing guide, which we obtain from our partner company.

The provided data further includes transactions for all of the client shippers' TL loads that originate and terminate within the continental United States, each loads' corresponding carrier tender sequence beginning with the primary carrier through any backup carriers if needed and each carriers' acceptance or rejection decisions until a carrier finally accepts the load.

After data cleaning and pre-processing, we retain just under four years of transactions; from September 2015 to May 2019. In particular, the data spans one complete market cycle into the next - that is, a soft period, a tight period, and the beginning of a second soft period, which are defined and tested in {§}4.2. For consistency, we consider only TL, dry van, long haul (greater than 250 miles) loads, since pricing schemes differ for other freight types and shorter lengths of haul. As a result, we maintain 1,933,299 unique loads offered by 71 shippers, tendered to 1,650 primary carriers. They comprise 7,573 shipper-primary carrier pairs operating from 4,915 origin 5-digit zip codes to 9,668 destination 5-digit zip codes. The dataset also contains geographic regions defined by our industry partner as key economic zones of shippers' business. The continental United States is divided into a set of 135 mutually exclusive and collectively exhaustive regions. Each of these regions represents an origin and a destination region.

Table 1 below summarizes the data set. In addition to these variables, the metrics described in the following subsections are computed to measure attributes of the S-C relationship.

\begin{small}
\begin{table*}[ht]
\caption{Data Summary}
\centering
\begin{tabular}[c]{ |c| p{7cm}| } \hline
Date range & Sept 1, 2015 - May 30, 2019 \\ \hline
Shipment Type & TL, dry van, long haul ($\geq$250 mi) \\ \hline
Num. Loads & 1,933,299 \\ \hline
Num. Shippers & 71 \\\hline
Industries & Automotive, Food \& Beverage or Consumer Package Goods,  Manufacturing, Paper \& Packaging, Other\\\hline
Num. Primary Carriers & 1,640\\
 & (Asset-based: 1,446, Non-asset: 194) \\\hline
Num. Shipper-(Primary)Carrier pairs & 7,573 \\\hline
Num. Origin 5-digit Zips (Regions) & 4,915 (135) \\\hline
Num. Destination 5-digit Zips (Regions) & 9,668 (135) \\\hline
\end{tabular}
\end{table*}
\end{small}

\subsection{Break Point Detection and Market Period Definition}
First, we define the market periods by detecting breaks in  market data. The time series of aggregated weekly PAR across all shippers is used as a proxy for market conditions. Previous literature (\cite{Scott2017}, \cite{Scott2015}, and \cite{Kirk2013}, for example), use monthly spot market prices to characterize market conditions. In our dataset only about 2\% of the total transactions go to the spot market and this value changes with market conditions as more loads go to the spot market during tight periods, As such, our data is limited in using spot prices to define market periods as other studies have.

However, \cite{Scott2017} show a 77.3\% correlation between spot premium time series data and the Morgan Stanley supply and demand sentiment index, which surveys a broad array of shippers, carriers, and brokers. Our weekly national PAR series and this same Morgan Stanley weekly freight index from 2015 to 2019 have a correlation of -85.2\%. That is, we show that high demand for capacity correlates with lower PAR - the definition of a tight market. Not only are supply and demand over time strongly correlated with spot prices as demonstrated in \citet{Scott2017}, we show that they are strongly correlated to our measure of the market, PAR. Thus we can be confident in choosing PAR rather than other metrics (e.g., spot prices, freight industry indices such as national load-to-truck ratio, truck manufacturing order rates, or other macroeconomic indicators) to define market conditions; acceptance ratios capture real-time carrier decisions.

In practice, many shipper monitor the changing market conditions by tracking the PAR along with outer routing guide failures – not just spot market price changes. In fact, one could argue that PAR is a more sensitive metric than spot premium as it captures contract abandonment by carriers. Thus, PAR is an appropriately representative and sensitive measure of how carriers respond to and communicate with shippers and, thus, of overall market conditions. Further, by considering the aggregate measure, the break points detected do not presuppose this study's results. 

To define and validate the identified breaks, we combine two distinct break point detection methods. As neither method perfectly captures the changes we expect, we run each method independently and take points that are identified by both methods as validated breaks. This results in a more robust break point detection method.

\subsubsection{Quandt Algorithm }
The Quandt method, developed by \cite{Chow1960} and expanded by \cite{Quandt1960} is the standard econometric test for a break point that is not known a priori. The method calculates a sequence of Pearson's chi-square statistics to determine when the series has shifted \citep{Hansen2001}. The algorithm considers each point in the time series as a candidate break point, $\tau$. The data is split into sub periods at $\tau$: $A$ for $t\leq\tau$, and $B$ for $t>\tau$. We estimate the vector of coefficients of a linear regression fit to each sub period and to the full series without a break (denoted by T):

\begin{align}
  \hat{y}_t^A &= \hat{\beta}_0^A + \hat{\beta}_1^AX_t^A + \varepsilon^A, \quad \quad t\leq\tau \nonumber\\
  \hat{y}_t^B &= \hat{\beta}_0^B + \hat{\beta}_1^BX_t^B + \varepsilon^B, \quad \quad t>\tau\\
  \hat{y}_t^{T} &= \hat{\beta}_0^{T} + \hat{\beta}_1^{T}X_t^{T} + \varepsilon^{T}, \quad \quad \forall t \nonumber
\end{align}

The chi-square distribution is used to assess the statistical significance of the difference between the two regression models and the equality of the coefficients is tested using an $F$-statistic (the ratio of two chi-square distributions) \citep{Chow1960,Hansen2001}.

The break point, $\hat{\tau}$, is determined to be the point at which the maximum $F_{\tau}$ is achieved: $\hat{\tau} = {argmax}_{\tau}{F_{\tau}}$

Typical $F$-statistic critical values for such tests are not appropriate for $F_{\hat{\tau}}$ when the break point is not known a priori and the Quandt method is applied \citep{Hansen2001}. However, critical value tables for the Quandt statistic and $p$-value calculation methods have been developed by \cite{Andrews1993}, \cite{Andrews1994}, and \cite{Hansen1997}.

The Quandt algorithm described above considers all points in the time series for which $\tau \in [\tau_0, \tau_1]$, where we exclude the tails of the data from consideration as candidate points to maintain enough observations on either side to estimate meaningful regression models \citep{Quandt1960,Hansen2001}. The sample is trimmed to the interior $\lambda$ to $(1-\lambda)$ range of the data. \cite{Hansen2001} suggests that typical values of the trimming parameter, $\lambda$, fall between 0.05 and 0.15. In this study we use $\lambda = 0.05$ trim and address its sensitivity by defining a transition period on either side of a detected break point when defining our market periods.

\subsubsection{Considerations of the Quandt Algorithm } While well documented in econometric and statistics literature, the Chow and Quandt algorithms have a number of drawbacks when applied to the data presented in this paper. First, the methods are well suited for time series data containing a single break, and despite extensions developed to account for multiple breaks, the results of these extensions are sensitive to both the date range and the trim parameter selected.

The second drawback of the Quandt algorithm as applied to this research is that it tests for locations in which one can fit the best linear model to each segment of the data. However, that is not the intent of this step in our research. We merely require a method to determine when average values of the aggregated PAR has changed. Unfortunately, break point detection based on average values is not found extensively in the literature. As such, we combine the method described above with a sequential difference of means test described in the following subsection.

\subsubsection{Jumping Mean Approach}
We are interested in determining the times at which the average value of the aggregated PAR changes. To identify these points, for each date in the time series, $\tau$, we compute the average of the PAR values for the preceding $W$ weeks and the following $W$ weeks. These backward- and forward-looking average values are denoted by $\overline{X}^{b,W}_{\tau}$ and $\overline{X}^{f,W}_{\tau}$, respectively and we take the absolute value of the difference between the two values, $D_{\tau}^W$:

\begin{equation}
    D_{\tau}^W = \lvert \overline{X}^{b,W}_{\tau} - \overline{X}^{f,W}_{\tau} \rvert
\end{equation}

Where
\begin{equation}
\begin{gathered}
    \overline{X}^{b,W}_{\tau} = \dfrac{1}{W}{\sum_{i=0}^{W-1} X_{\tau-i}}\\
    \overline{X}^{f,W}_{\tau} = \dfrac{1}{W}{\sum_{i=1}^{W} X_{\tau+i}}\\
\end{gathered}
\end{equation}

In doing so, we draw from a data mining context, which describes such a `jumping mean' break point measure, where a break is detected if a difference score (which we define here as the absolute difference in means between the forward- and backward-looking window of the data) is large enough \citep{Tak2016}. We consider the points at which the local maxima of the absolute difference occur as potential break locations, as suggested by \cite{Hansen2001} and \cite{Bai1997} for the Quandt method.

Next, applying a standard $t$-test, we check the statistical significance of the difference of means at the potential breaks, $\tau$. If $t_{stat} > t_{crit}$ at the desired significance level we consider it as a potential break point.

We test the sensitivity of this combined method to the window size, $W$. Windows of 6, 10, 12, 26, 34, and 52 weeks are tested and we determine that a satisfactory window size is 12 weeks, as smaller windows are too sensitive to variability in the data, detecting many immaterial break points, and larger windows over-smooth, missing potentially important breaks (see Appendix A for window sensitivity test results).

Next, we combine the results of the Quandt algorithm and this jumping mean approach. We retain only break dates that are identified by both the Quandt method and the jumping mean approach with a 12-week window. These dates are summarized in Table 2. The resulting validated break points are the first week of February 2016, the first week of July 2017, and the second week of January 2019 (see Figure 1).

\begin{small}
\begin{table}[h!]
\caption{Break points identified by the Quandt method and jumping mean approach}
\centering
\begin{tabular}{c|c} \hline
Quandt results & Jumping mean\\ \hline \hline
02-05-16 & 02-07-16 \\ \hline
 & 05-29-16 \\ \hline
 & 01-29-17 \\ \hline
06-28-17 & 06-25-17 \\ \hline
 & 09-02-18 \\ \hline
01-16-19 & 01-20-19 \\ \hline
\end{tabular}
\end{table}
\end{small}

\subsubsection{Market Period Definitions }
While determining a break at a single point is simple and convenient, claiming that an economic market change takes place at a single specified date is not realistic. We use the break point detection method described above as a first order approximation of a structural change point \citep{Hansen2001}. To account for seasonal effects, we take a full year of data for each defined time period. By selecting a full year during each market period we average out any seasonality individual shippers or industries observe. In this way, seasonality in demand for distinct shippers or industries in one market period is compared to the same seasonality in the second market period and allows us to compare across shippers of different size and industry. Further, in order to allow for market transition buffer time, we choose the year of data that is centered between our validated break dates.

While it is reasonable to consider economic markets as continuous rather than discrete as we do here, we define our research question such that we measure market-specific behaviors, and thus must define market periods. To procure transportation services, shippers typically run annual national bid events, which are essentially seen as step changes resulting in rates and primary carrier agreements that are expected to be in effect for a year. These step change, which represent how the shipper is responding to the current market conditions, occur are different times throughout the year for different shippers. The result of cumulative shipper annual bids eventually leads to a distinct market type. Moreover, the decision to discretize the market is based on conversations with industry practitioners, who tend to refer to distinct ``soft'' and ``tight'' market periods. Our resulting market periods are defined in Table 3 below.

\begin{small}
\begin{table}[h!]
\caption{Market Periods}
\centering
\begin{tabular}{c||c|c} \hline
 & Soft & Tight\\ \hline \hline
Date Range & Apr. 15, 2016 – Apr. 14, 2017 & Oct. 1, 2017 – Sept. 30, 2018 \\ \hline
Avg. Weekly PAR & 81.9\% & 68.5\% \\ \hline
Avg. Weekly CPL & \$1155.9 & \$1420.7 \\ \hline
\end{tabular}
\end{table}
\end{small}

Figure 1 shows both the average weekly aggregated PAR and the average weekly accepted cost per load (CPL) over time overlaid with our identified break points and market periods. This CPL is the linehaul price that is accepted, which includes primary carriers, backup carriers within the routing guide, and spot loads. The two measures mirror each other as expected: as the market tightens, acceptance ratios decrease and prices paid increase. As the market softens again, acceptance ratios increase and accepted prices decline.

\smallbreak

\begin{figure*}[ht]
\begin{center}
  \caption{Validated Break Dates, Market Periods, Aggregated Primary Carrier Acceptance Ratio (PAR) (left axis), and Accepted Cost Per Load (right axis)}
  \includegraphics[width=0.9\textwidth]{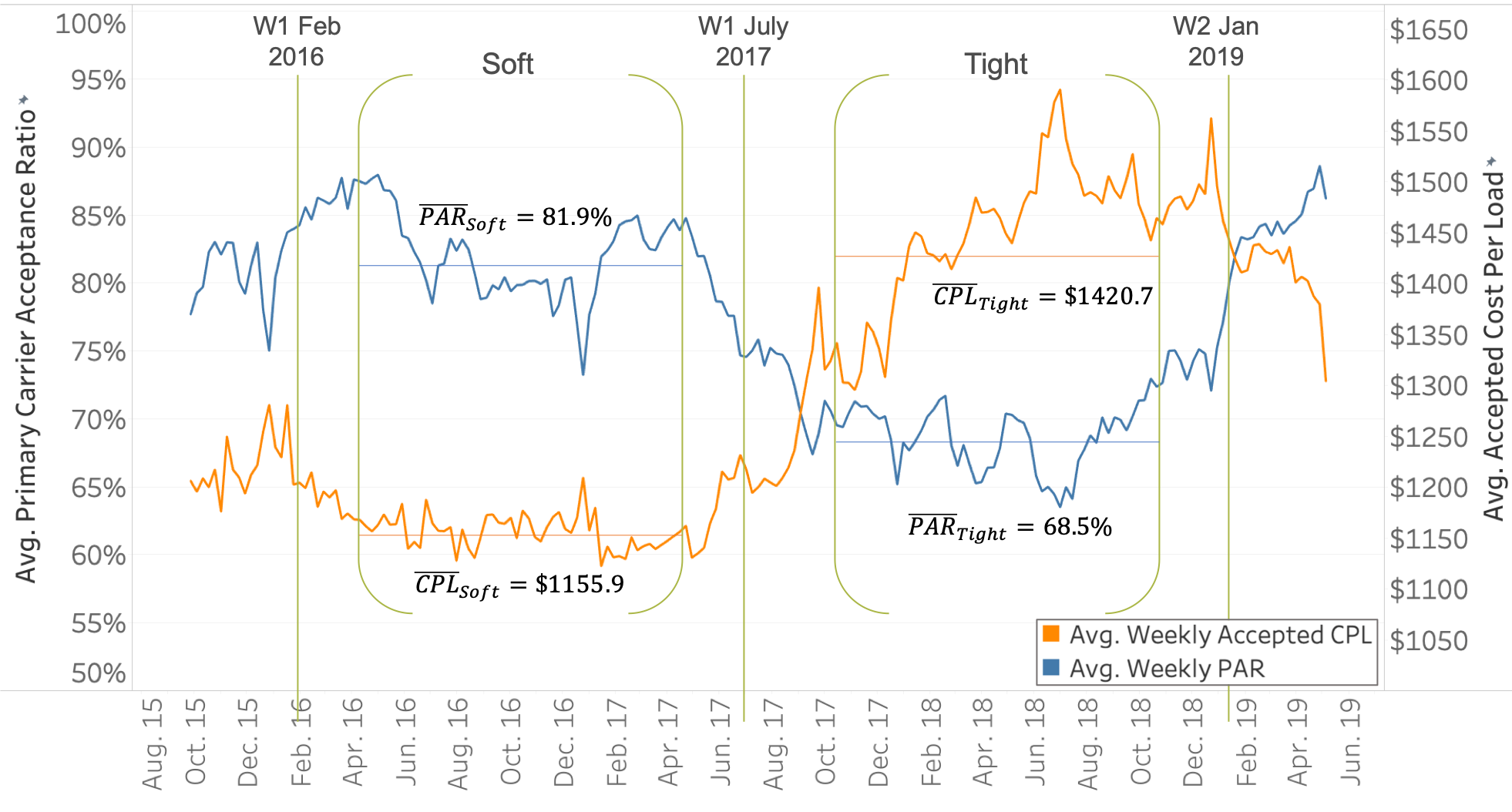}
\end{center}
\end{figure*}

\subsection{S-C Relationship Measures}
We select and define the S-C relationship metrics based on a combination of relationship attributes studied in the inter-firm relationship literature and in TL transportation literature, TL transportation industry reports, and conversations with our industry partner and with professionals at both shipper and carrier organizations. Thus, we have developed a set of relationship characteristics based in both theory and practice.

We define the relationship between shippers and carriers along six shipper behaviors measured in each of the market periods, The first, Primary carrier Acceptance Ratio (PAR), in the tight market is our dependent variable in our model. The remaining set of shipper behaviors is composed of pricing relative to market rates, volatility of volume offered measured, tender lead time, and origin and destination dwell times. In addition, we include one characteristic of S-C interactions across time periods (cadence of tenders), two shipper characteristics (size and industry), and two carrier characteristics (fleet size and asset- versus non-asset). We detail these measures in the following subsections and report summary statistics in Appendix C.

\subsubsection{Primary carrier Acceptance Ratio }
Primary carrier Acceptance Ratio (PAR)\footnote{Primary carrier AR (PAR) is a preferred measure to AR across all carriers, where the same load may be rejected multiple time.} is measured as the weekly fraction of loads that are accepted by the primary contract carrier relative to the total number of loads offered to that carrier from a specific shipper:

\begin{equation}
    PAR^{a,b}_{i,j,w} = \dfrac{L^{a,b}_{i,j,w,accepted}}{L^{a,b}_{i,j,w, offered}}
\end{equation}

where $L$ is the number of loads either accepted or offered; $a$ and $b$ denote the shipper and carrier, respectively; $i$ and $j$ denote the origin and destination regions, respectively; and $w$ is the week in which loads are offered.

High PAR in a soft period does not necessarily indicate a good S-C relationship because demand is low in soft markets and carriers search for business regardless of their relationship with shippers. On the other hand, low PAR in a soft period is a signal of a poor S-C relationship, as it is in the portion of the market cycle when carriers need loads. On aggregate, PAR decreases as the market turns from soft to tight, by definition of the market types. However, we expect that S-C pairs with an existing strong relationship will not experience a decrease in PAR as severe as that of the aggregate market PAR decline.

We calculate PAR for a S-C pair on each of their lanes, and in each week in which loads are offered between them. This value is averaged over all weeks in each of the market periods and we obtain an average weekly acceptance ratio, which is averaged again across all lanes between a shipper and carrier in each market period to obtain the S-C PAR for both periods: $PAR^{a,b}_{soft}$ and $PAR^{a,b}_{tight}$\footnote{We do not weight the PAR by volume offered, number of weeks in the year loads are offered, or number of lanes between the S-C pair. As a result, the relative importance of each load is higher for S-C pairs with less total volume (or fewer weeks or lanes) than that of S-C pairs with more volume (or weeks or lanes).
}.

\subsubsection{Market Rate Differential }
Market Rate Differential (MRD), or the percent above or below market benchmark prices at which a load moves, measures how much of a premium (or discount) a shipper pays its primary carrier. As noted earlier, we expect that shippers paying high soft period prices, when the market is in their favor, will receive better performance in the next tight period - that is, carriers reciprocate high soft period pricing with high tight period PAR. On the other hand, we expect shippers that pay low soft market prices are taking advantage of their position and carriers respond with decreased PAR in the following tight period.

We determine MRD by calculating a benchmark price for each lane against which the S-C pairs' contract prices are to be compared. Our load data is split into the soft and tight periods to obtain lane benchmark rates for each time period. Freight transportation costs can be reduced to a linear combination of fixed costs and variable costs associated with a distance measure \citep{Daganzo2004}. The benchmark rate captures the regional nature of TL pricing with lane-specific fixed effects by including origin and destination indicator variables. In doing so, we incorporate effects of outbound loads from the origin and inbound loads to the destination. These factors are important as carriers makes load acceptance decisions, in large part based on where they have available capacity, where they will ultimately end up, and the likelihood of obtaining follow-on loads.

We implement a multiple linear regression model with heteroskedastic robust standard errors in which the load linehaul price is regressed on an origin region indicator variable, a destination region indicator variable, and distance variable \citep{Ballou1991,Scott2015}.

Specifically, an indicator variable is created for all but one of the origin regions and all but one of the destination regions. The lane which corresponds to this origin-destination pair is the base case lane. The base case origin and destination regions are chosen based on their outbound and inbound volume, respectively. We measure the total outbound (inbound) volume in both of the time periods, $t \in \{Soft, Tight\}$, for each region. The base case origin (destination) region is chosen if it is ranked in the top five regions by volume in both time periods. If more than one region qualifies, we take the region with the highest cumulative volume. We choose a high-demand lane as our base case to ensure it is representative of aggregate market conditions.

Our resulting base case origin region is the greater Greenville, SC area and the base case destination region is the Dallas, TX metro region. We test this choice of base case origin and destination regions among top-ranked regions by volume and find that the differences between lane benchmark values relative to one another for varying base case regions do not impact statistical results.

The vector of coefficients that results from our linear regression includes an intercept term, which is the fixed transportation cost, $\hat{\beta}^t_{base}$, a distance (i.e. variable cost) coefficient, $\hat{\beta}^t_{dist}$, $I$-1 origin coefficients (where $I$ is the set of origin regions in the dataset), and $J$-1 destination coefficients (where $J$ is the set of destination regions in the dataset). These 135 key economic regions defined by our industry partner are geographic clusters of transportation demand patterns. The origin and destination coefficients of our linear regression model, $\hat{\beta}^t_i$ and $\hat{\beta}^t_j$, can be considered `price premiums', or the costs associated with an origin or destination different from the base case lane.

The origin and destination price premiums, $\hat{\beta}^t_i$ and $\hat{\beta}^t_j$, for each time period, are illustrated in Appendix B, as are $\hat{\beta}^t_{base}$ and $\hat{\beta}^t_{dist}$. In general, relative pricing of the regions hold through market periods. For example, in both the soft and tight markets, the Midwest and Southwest regions of the United States are typically higher priced regions to ship out of (i.e. positive and high origin price premiums). This is likely due to the large manufacturing hub in the Midwest the port authorities of Los Angeles and Long Beach causing high demand for outbound capacity. As a result, carriers can command higher prices. At the same time, the Northeastern region does not generate much freight volume relative to the rest of the country; carriers are willing to take lower prices in order to leave the region (i.e. negative origin price premiums in the Northeast), whereas higher prices are required to incentivize inbound carriers in the first place (demonstrated by high destination price premiums in the Northeast). Thus, these regional values define and quantify fore and backhaul lanes.

Each lane benchmark rate is calculated as follows. The base case lane benchmark price is the regression model intercept, $\hat{\beta}^t_{base}$, plus the distance coefficient, $\hat{\beta}^t_{dist}$, multiplied by the mean value of the distance variable of loads on that specific origin-destination pair, $\overline{X}_{dist}$. All other lane benchmark rates consist of the fixed cost intercept term, plus the distance coefficient and variable interaction, plus the origin and destination region coefficient price premiums corresponding to that lane, $\hat{\beta}^t_i$ and $\hat{\beta}^t_j$:

\begin{equation}
\hat{b}^{t}_{i,j} = \hat{\beta}^t_{base} + \hat{\beta}^t_{dist} \overline{X}_{dist}  + \sum_{i\in I, i\neq i_{base}} \hat{\beta}^t_i X_i + \sum_{j\in J, j\neq j_{base}}\hat{\beta}^t_j X_j
\end{equation}

where $X_i$ and $X_j$ are binary variables indicating which origin and destination is active for the lane of interest. Once the lane benchmark rates are established, the market rate differential is calculated per load, $k$, as the percent above or below the lane benchmark price. That is, the difference between the load linehaul price and the corresponding lane benchmark rate, relative to the benchmark rate:

\begin{equation}
MRD^t_{k,a,b} = \dfrac{LH^t_{k,a,b} - \hat{b}^{t}_{i,j}}{\hat{b}^{t}_{i,j}}
\end{equation}

We obtain the market rate differential for a S-C pair, $MRD^t_{a,b}$, by averaging over all loads on each lane between the shipper and carrier, and averaging across all of their lanes for each market period separately.

In the next subsections we discuss shippers' operational performance that may impact the S-C relationship and thus carrier reciprocity in tight market conditions. A natural measure may be the the total business offered to the carrier. However, the total amount of volume a shipper tenders to a carrier does not necessarily lead to better carrier performance, as the business offered must fit the carrier’s network. For example, previous literature on the benefits of information sharing finds two seemingly opposing results. In a dyadic supply chain, \cite{Helper2010} demonstrate that supplier capacity limits the value of sharing information. However, \cite{Bakal2011} show that the benefit of information sharing is lower with suppliers with less capacity. As such, we use consistency (or CV, volatility), tender lead time, offer cadence as operational measures of good shipper tendering behavior rather than pure volume tendered.

\subsubsection{Tendered volume volatility }
TL carriers commonly call for consistency of demand at the load and lane levels. For the shipper, this means minimizing variation in tendering behaviors. Industry reports from J.B. Hunt, one of the largest TL carriers in the US, and C.H. Robinson, a leading third party logistics (3PL) provider in the US, cite increased consistency as an action shippers can take to improve their relationships with carriers, improve freight acceptance, reduce cost per load, and allow carriers and their drivers to better optimize profitability (\cite{JBHunt} and \cite{CHR2015}). Further, previous studies find consistency measures to be significant indicators of spot market load acceptance \citep{Scott2017}, reduced cost per load \citep{Harding2005}, and reduced routing guide failure \citep{Yuan2019}.

We measure variability of volume tendered by a shipper as the coefficient of variation (CV) of the weekly volume offered. This normalized measure indicates what percent of the mean the standard variation is. It is used rather than raw standard deviation in order to compare the variability measure across dissimilar lanes. A shipper with high tendered volume variability is expected to have low PAR because inconsistency causes difficulty for carriers' network planning. As a result of high variability, either trucks will not be available when and where shippers need them, or carriers will not make an effort to ensure these inconsistent shippers are served. Thus, we expect that higher volatility from shippers leads to lower PAR in tight markets.

\subsubsection{Tender lead time }
In order for carriers to balance capacity utilization with availability, they need enough time to respond to the tender requests, ensure a truck is available, and reposition it to the pickup location on time. According to J.B. Hunt, in addition to consistency and predictability, carriers seek reasonable lead times, which allow them to create a schedule and optimize driver’s hours \citep{JBHunt}. For example, \cite{Caldwell2008} found that with greater tender lead time, a shipper saw lower variability in prices paid for loads. We measure this tender lead time (TLT) as the number of days between when the load is first offered to the primary carrier to when it needs to be picked up.

It is important to note however, that with too many days of advanced notice, much can change in that time. For example, the appointment time may need to change, the carrier may not have capacity available as expected because of previous service delays, or the shipper may have been tendering the load to gather information of general carrier availability or willingness to serve for the tendered price. This point is illustrated by the numerical results of \cite{Zolfagharinia2014}, which find that with a two-day TLT, the carrier’s profit increases 22\% over the base case of one day, but three days of advanced load notice only increases the carrier's profit by 6\% over the base case. \cite{Lindsey2015} and \cite{Scott2015} study the impact of lead time on spot prices. The former consider a dummy variable for loads with greater than 8 days of lead time, while the latter finds that the impact of lead time drops off quickly for TLT longer than two days. \cite{Tjokroamidjojo2006} model the benefit of advanced load information sharing (i.e., TLT) to minimize carrier’s total cost and find that carriers seek longer TLT. However they do not address the potential non-linearity in benefit from increased TLT addressed by some of the other studies.

\subsubsection{Origin and destination dwell time }
Carriers are often concerned with delays that occur during pickup and delivery. In a study of shipper behaviors that impact carrier performance \cite{CHR2015} finds that carriers cited dwell times in their top shipper characteristics important to price and service decisions. Dwell time becomes even more important when one considers the regulations drivers face in terms of hours of service (HOS) laws\footnote{In essence, a driver has 14 hours of “on-duty” time with a required 30-minute break. 11 of those 14 hours can be spent driving and 2.5 hours can be spent on all other activities, including pickup, delivery, safety inspections, and shutdown. However, it is not uncommon for live loading and unloading to take longer than 2.5 hours. When these activities do, it eats into the drivers’ precious 11 hours of ``on-duty, driving'' time they could otherwise be using to make a profit.}. Particularly in times when capacity is tight and carriers want to retain drivers, drivers’ time utilization goes hand in hand with asset utilization and thus, carriers’ attitudes toward shippers \cite{JBHunt}.

Dwell time is also considered in the literature; \cite{Zolfagharinia2014} and \cite{Tjokroamidjojo2006} include dwell time in their modelling approaches as a cost incurred by carriers. 

Delays during delivery can often be more problematic than pickup because drivers may be able to make up some of that added time during the move. However, delays at the destination may make the driver late to the next job's appointment, which may then add further delays to that pickup, thus, proliferating the problem. We expect then, that destination dwell times may have higher impact on PAR than that at origins.

\subsubsection{Offer cadence }
In line with carriers' desire for consistency and the ability to plan for demand, the frequency, or cadence, at which shippers tender loads is often attributed to carrier's freight acceptance. In a general supplier-customer setting, \cite{Rinehart2004} study inter-firm behaviors and argue that interaction frequency is a key success factor to supplier-customer relationships. \cite{Scott2017} includes the number of days since the previous load was offered to a spot carrier by a shipper as a measure of load offer frequency. Interviews with practitioners and industry reports by J.B. Hunt and C.H. Robinson demonstrate that tender cadence, as measured by weeks in which loads are offered to the carrier, is a measure carriers use to assess shipper relationship (\cite{JBHunt} and \cite{CHR2015}.

We measure frequency at which a shipper offers loads to a carrier as the number of weeks in a year in which the shipper offers at least one load to its primary carrier. We compare a metric, offer cadence, or the percentage of weeks during the year in which the shipper tenders loads to its primary carrier. For individual S-C pairs in our dataset, this percentage for the soft period and the tight period are highly correlated (with correlation coefficient of 0.90) so we use an average measure across the two market periods. Shippers that offer loads more frequently are expected to receive higher PAR, as the carriers rely on frequent loads to move trucks around their network and ensure trucks are available in the right location and at the right time for the portfolio of shippers served.

Table 4 summarizes the S-C attributes we consider and the literature and the academic and industry reports that have previously considered such attributes. This is not meant to be a comprehensive list, but a subset of the literature most closely related to our research question at hand.

\begin{small}
\begin{table*}[h!]
\caption{S-C Relationship Metrics in the Literature}
\centering
\begin{tabular}{c|c|c|c|c|c|c} \hline
Study & PAR & Pricing & Volatility & TLT & Dwell & Cadence \\ \hline \hline
\cite{Yuan2019}  &\checkmark&\checkmark&&&&\checkmark\\\hline
\cite{Caldwell2008} &&\checkmark&&\checkmark&&\\ \hline
\cite{CHR2015} &\checkmark&\checkmark&&\checkmark&\checkmark&\\ \hline
\cite{Harding2005} &\checkmark&\checkmark&\checkmark&&&\\ \hline
\cite{JBHunt} &&&\checkmark&\checkmark&\checkmark&\checkmark\\ \hline
\cite{Kim2013} &\checkmark&\checkmark&\checkmark&&&\\ \hline
\cite{Lindsey2015} &&&&\checkmark&&\\\hline
\cite{Rinehart2004} &&&&&&\checkmark\\ \hline
\cite{Scott2015} &&\checkmark&&\checkmark&&\\ \hline
\cite{Scott2017} &\checkmark&\checkmark&\checkmark&&&\checkmark\\ \hline
\cite{Tjokroamidjojo2006}&&&&\checkmark&\checkmark&\\ \hline
\cite{Zolfagharinia2014}  &&&&\checkmark&\checkmark&\\\hline
\cite{Zsidisin2007} &\checkmark&&&&&\\ \hline
\end{tabular}
\end{table*}
\end{small}

Finally, as the main focus of our study is to examine the impact of shipper behaviors and market conditions on tight market PAR, we control the impact of other factors that characterize the shippers and the carriers, described below.

\subsubsection{Shipper and carrier characteristics }
In this section, we describe the attributes of the shippers and carriers themselves that we include in our model to control for fixed effects. This allows us to determine if certain types of shippers or carriers behave or experience impacts from market changes differently from others. These fixed effects include shipper size, shipper industry, carrier size, and carrier service type.\\

\textbf{Shipper size: } Previous S-C relationship literature has been limited in the scope of available shipper data, however our dataset consists of hundreds of shippers. Thus, we contribute to the literature by considering the impacts of shipper characteristics. First, we consider the shipper's size, which we measure as the log (base 10) of average total annual volume tendered to all primary carriers. Larger shippers may be better insulated from the impacts of the market and maintain better carrier relationships. On the other hand, smaller shippers may invest in relationships with a smaller core set of carriers, and thus have better carrier relationships. We test if a shipper's sizes impacts how it experiences carrier performance as the markets shift.\\

\textbf{Shipper industry: } We similarly utilize the granularity of our available data by including the shipper's industry vertical in our model. Each shipper falls into only one of six industry categories: automotive, food \& beverage or consumer package goods (F\&B/CPG), manufacturing, paper and packaging, or a final catch-all category, other.\\

\textbf{Carrier fleet size: } Carriers with different attributes operate - and thus make decisions - differently. We first consider asset-based carriers' fleet size, measured as the log (base 10) of the carrier's tractor count, as an indicator of tight market PAR. For smaller carriers such as the single truck owner-operator, or those with only a few trucks, when demand is very high as is seen in tight markets, accepting tendered volume becomes more difficult than for larger carriers, as these small carriers have fewer available trucks to cope with the additional demand.\\

\textbf{Carrier service type } The second carrier attribute we consider is the carrier's service type. We split our data into S-C pairs for which the primary carrier is an asset-based carrier, which can be further segmented by fleet size as describe above, and a second set for which the primary carrier is a non-asset provider such as a brokerage. These providers do not own trucks, yet they may be contracted with shippers to provide services in a similar way to the asset carriers. Instead, they match their contracted shippers' needs with a vast pool of available capacity composed of many carriers. Brokerages utilize their extensive network of disagragate capacity to tailor the shipper-carrier match to the needs of both parties.

We are aware of one study that considers the differences in behaviors between asset and non-asset carriers, specifically, on spot load pricing strategies \citep{Scott2018}. The author uses auction theory to address the differences in how asset and non-asset carriers consider the two main decisions made in auctions - whether and how much to bid. He finds that non-asset carriers bid more frequently and higher than their asset counterparts. This study, one of the few empirical studies of actual firms' decisions repeated auctions over time as the author claims, underscores our claim that asset-based carriers and non-asset carriers behave differently and justifies our consideration of them separately.

In our dataset, the non-asset carrier for these S-C pairs is defined as the brokerage itself, not the carrier that ultimately matched and moved the load. That is, we do not have visibility on which trucking company actually moved the load. Thus, we can characterize the shipper-third party relationship by including non-asset carriers separately from shipper and asset-based carrier pairs. Our dataset of S-C pairs contains 159 of these shipper-non-asset primary carrier pairs. We include the non-asset carriers in a separate model (i.e., models 1b and 2b).

\subsection{Model Specification}
As we are interested in studying the S-C relationship between different markets, our unit of analysis is at the S-C pair level and our dependent variable is tight market PAR.

As this dependent variable is a continuous fraction that can take values between and including 0 and 1 – that is, if the primary carrier accepts none or all of the tendered volume from that shipper in the market period, respectively - and we directly observe this outcome, we implement a generalized linear model. Specifically, we use a beta regression \citep{Ospina2012}. We choose a logistic link function using a maximum likelihood estimation method and robust standard errors to allow for misspecification of the prior distribution, as discussed in \cite{Papke1996}, \cite{Figueroa2013} and \cite{Pereira2014}.

Beta regression uses the beta distribution as the likelihood for the dependent variable: 

\begin{equation}
    f(y_i\,|\,a,b) = \frac{y_i^{a-1}(1-y_i)^{b-1}}{\textrm{B}(a,b)}
\end{equation}

Where $\textrm{B}(\cdot)$ is the beta function defined by
\begin{equation}
    \textrm{B}(a,b) = \frac{\Gamma(a)\Gamma(b)}{\Gamma(a+b)}
\end{equation}

$a$ and $b$ are the shape parameters, and $\Gamma$ is the Gamma function \citep{Ferrari2004}.

We choose $a=4$, $b=1$, which fits our tight market PAR distribution with a coefficient of determination of the Q-Q plot of our tight market PAR and beta distribution of 0.931. The maximum likelihood estimators are calculated over $a$ and $b$ numerically through an iterative fitting process (see \cite{Nelder1972} and \cite{Little2013}). 

The beta regression with logistic link function allows us to perform a logistic transformation of the bounded dependent variable, transforming it to the real number line, and still retain extreme values (i.e., 0 and 1) \citep{Ferrari2004}. The logistic link model is defined as follows:

\begin{equation}
\begin{gathered}
    logit(y_i) = log\Big(\dfrac{y_i}{1-y_i}\Big)=x_i^T\boldsymbol{\beta}\\
    y_i \sim \textrm{Beta}(a,b)
\end{gathered}
\end{equation}

The regression coefficients, $\boldsymbol{\beta}$, are interpreted as the log odds ratio of the dependent variable, tight market PAR, for each independent variable, $x_i$. We can transform the coefficients back to actual tight market PAR by exponentiating eq. (9) and solving for $y_i$.

As we are testing four hypotheses, we develop four models. All four models include a categorical variable for shipper industry vertical, and omit Manufacturing as our baseline. Models 1a and 2a test whether asset-based primary carriers consider previous soft period shipper behaviors in tight market period freight acceptance decisions (H1a) and whether they consider current market tight market period shipper behavior - that is, if they are myopic (H2a), respectively. These models include the continuous variable, carrier fleet size, as a predictor of tight market period PAR.

Models 1b and 2b correspond to hypotheses H1b and H2b, respectively. As such, rather than carrier fleet size, these models include a binary variable indicating whether the carrier is asset-based - coded as a 1 - or not - coded as a 0. This means that the reported coefficient for the asset binary variable in these models is associated with the additional (log odds ratio of) tight market PAR that would result if the carrier is asset-based (i.e., $x_{asset} = 1$ rather than $0$).

\section{Results}
In this section, we first discuss whether carriers have long memories; that is, the results of the beta regression model of tight period PAR on soft market shipper behaviors (i.e., models 1a and 1b). We then discuss whether carriers are responding to current tight market shipper behaviors with the results of models 2a and 2b.

\subsection{Carriers' responses to \textbf{previous} shipper behavior}
Table 5 reports the results of the beta regression analysis of soft market shipper behaviors on tight market PAR for asset-based primary carriers and for all primary service provider types (models 1a and 1b for H1: Do carriers remember shippers' previous soft market behavior?).

Hypothesis 1a is not supported, and except for one soft period shipper behavior, hypothesis 1b is also not supported. This result means that in general, shippers’ soft market period behaviors do not significantly impact their primary carriers’ tight market period PAR. However, soft market PAR (which is not a shipper behavior, but rather an indicator of previous \textit{carrier} behavior) is a predictor of tight period PAR. This suggests that on average, primary carriers tend to maintain their acceptance performance across markets.

The results of model 1b suggest that for non-asset primary carriers, destination dwell time in the soft period as a significant predictor of tight period PAR. This finding, that non-asset service providers may be more sensitive to soft market period destination dwell time than asset-based carriers, is explained by our industry partner. For asset-based carriers, one driver may handle the long-haul move to get the load into the destination region, drop the trailer at a regional facility or distribution center, and move on to the next load. A regional driver will then take the loaded trailer to the final destination. In soft markets, asset utilization is low and asset carriers may have the flexibility needed to have multiple trucks handle a single load than their non-asset counterparts, which have less control over shuffling drivers and trucks around. This may explain the greater sensitivity to destination dwell time in the soft market observed when non-asset carriers are included in the analysis of models 1b as compared to model 1a.

A number of control factors that describe the S-C interaction across market periods, shipper-specific characteristics, and carrier-specific characteristics are statistically significant in models 1a and 1b. These factors are also significant in models 2a and 2b, and we discuss them further in section 5.2, using coefficients found from model 2a.

The main result of H1 is that primary carriers do no remember how shippers treated them in the past, when markets favored the shippers and they may have acted opportunistically. Put another way, shippers do not benefit from higher PAR in tight markets for having good performance or competitive pricing in previous soft markets.

\begin{table}[htb!]
    \centering
    \caption{Model 1 results, soft market shipper behaviors}
    \begin{tabular}{l|c|c}
        Variable & \shortstack{Model 1a\\ Asset-based \\ primary carriers} & \shortstack{Model 1b\\ Non-asset \\ primary carriers}\\\hline\hline
        Constant & -0.4818 & 2.681 \\
         & (0.6803) & (1.209) \\\hline
        PAR$_{soft}$ & 2.630*** & 1.275** \\
         & (0.3574) & (0.6828) \\\hline
        MRD$_{soft}$ & 0.0067 & 0.0111 \\
         & (0.0040) & (0.0091) \\\hline
        CV$_{soft}$ & 0.0334 & -0.1092 \\
         & (0.2708) & (0.4889) \\\hline
        TLT$_{soft}$ & -0.0255 & -0.0251 \\
         & (0.0222) & (0.0589) \\\hline
        Origin dwell$_{soft}$ & 0.0288 & -0.0606 \\
         & (0.0350) & (0.1515)\\\hline
        Destination dwell$_{soft}$ & -0.0664 & -0.3592**\\
         & (0.0563) & (0.1546)\\\hline
        Offer cadence & 0.8143*** & 0.4902 \\
         & (0.2421) & (0.4515)\\\hline
        Log shipper volume & -0.1462** & -0.1692*\\
         & (0.0630) & (0.1029)\\\hline
        Automotive & 0.9791*** & -1.057**\\
         & (0.2370) & (0.5292)\\\hline
        F\&B/CPG & 0.2416 & -0.8392*\\
         & (0.1839) & (0.4945)\\\hline
        Paper \& Packaging & 0.0261 &-1.330***\\
         & (0.2084) & (0.5355)\\\hline
        Other & 0.3821 & -1.096\\
         & (0.3851) & (0.7793)\\\hline
        Log carrier fleet size & -0.0449 & NA \\
         & (0.0327) & NA \\\hline
    \end{tabular}
    \smallbreak
    Note: robust standard errors reported in parentheses \\
    significance level: *0.1; **0.05; ***0.01
\end{table}

\subsection{Carriers' responses to \textbf{current} shipper behavior}
Next, we discuss the results of models 2a and 2b, which test whether tight market shipper behaviors impact tight market PAR (H2).

The results, summarized in Table 6 demonstrate that H2a and H2b are supported: carriers respond to current tight market period shipper behaviors. In particular, tight period pricing (i.e., market rate differential), consistency and cadence of tendering patterns, and destination dwell times in tight market periods have significant impacts on PAR in the same tight market period.

\begin{table}[htb!]
    \centering
    \caption{Model 2 results, tight market shipper behaviors}
    \begin{tabular}{l|c|c}
        Variable & \shortstack{Model 2a\\ Asset-based \\ primary carriers} & \shortstack{Model 2b\\ Non-asset primary \\ carriers}\\\hline\hline
        Constant & 2.1214*** & 3.583***\\
         & (0.5835) & (0.1.113) \\\hline
        MRD$_{tight}$ & 0.0250*** & 0.0461*** \\
         & (0.0043) & (0.0088) \\\hline
        CV$_{tight}$ & -0.7876*** & -0.8715*\\
         & (0.2655) & (0.4885) \\\hline
        TLT$_{tight}$ & -0.0115 & 0.0989\\
         & (0.0216) & (0.0969) \\\hline
        Origin dwell$_{tight}$ & 0.0388 & -0.1997\\
         & (0.0327) & (0.1308)\\\hline
        Destination dwell$_{tight}$ & -0.1572*** & -0.2955**\\
         & (0.0492) & (0.1345)\\\hline
        Offer cadence & 1.0986*** & 1.144**\\
         & (0.2474) & (0.4922)\\\hline
        Log shipper volume & -0.1317** & -0.1433\\
         & (0.0633) & (0.1244)\\\hline
        Automotive & 1.0406*** & -1.002**\\
         & (0.2234) & (0.5167)\\\hline
        F\&B/CPG & 0.2750 & -0.9841*\\
         & (0.1821) & (0.5622)\\\hline
        Paper \& Packaging & -0.0350 & -1.288**\\
         & (0.2115) & (0.5275)\\\hline
        Other & 0.0550 & 0.7379***\\
         & (0.3719) & (0.7856)\\\hline
        Log carrier fleet size & -0.0523 & NA \\
         & (0.0310) & NA \\\hline
    \end{tabular}
    \smallbreak
    Note: robust standard errors reported in parentheses \\
    significance level: *0.1; **0.05; ***0.01
\end{table}

While the intent of this study is to indicate which shipper behaviors, S-C relationship measures, and shipper and carrier characteristics in different market periods impact tight market PAR, not to predict tight PAR, in this section we illustrate the relative impact of the significant variables with the models' resulting coefficients and offer some industry context. However, it is important to note that these models and their quantitative results have low predictive power.

As an illustration of model coefficients interpretation, the intercept term of model 2a, 2.1214, indicates that the model predicts tight period PAR to be 0.8932, or 89.32\% of tendered volume is accepted by the primary carrier. We obtain this result by plugging into eq.(9) and solving for PAR:

\begin{align*}
    log\Big(\dfrac{PAR_{tight}}{1-PAR_{tight}}\Big) = 2.1214
\end{align*}

Shippers that pay their primary carriers above tight market rates on average observe higher tight market PAR. Again, we plug in the $MRD_{tight}$ coefficient, 0.0250 with the constant coefficient, and solve for PAR:

\begin{align*}
    log\Big(\dfrac{PAR_{tight}}{1-PAR_{tight}}\Big) = 2.1214+0.0250
\end{align*}

which gives us $PAR_{tight} = 95.95\%$. This means that all else held equal, an increase of 1 percentage point in $MRD_{tight}$ results in an increase of 6.65 percentage points in $PAR_{tight}$. Discussions with shippers and carriers indicate that reasonable pricing changes may be in the range of 1-3\%. 

Next, we observe that shippers’ volatility in tight market period tendering behavior (i.e., CV of weekly offered volume) negatively impacts tight PAR. Recall that CV measures the ratio of standard deviation to mean of weekly tendered volume, as a percentage. Thus, one percentage point increase in tight market CV corresponds to an 11.3\% decrease in PAR (from 89.32\% to 79.19\%). This result is consistent with previous literature (\cite{Scott2017}, and \cite{Yuan2019}), which study load- and lane-level freight acceptance.

The results of model 2a further suggest that carriers are sensitive to tight period destination dwell time: an additional hour of tight market dwell time that a primary carrier experiences results in a 1.75\% decrease in tight market PAR (from 89.32\% to 87.73\%). Moreover, not only are shippers with high destination dwell times susceptible to reduced PAR, but they likely incur detention fees at facilities with excessive wait times. Our analysis, however, focuses on linehaul prices rather than these accessorial charges.

Some shipper and carrier characteristic are found to be statistically significant in all four models. Of these interaction factors across market periods, the cadence of tenders offered, is a significant contributor to tight market PAR. For example, using results from model 2a, a 10\% increase in offer cadence – which corresponds to 5.2 additional weeks per 52-week market period – corresponds to a 7.69\% increase in tight PAR (from 89.32\% to 96.17\%).

The results of all four models indicate that shipper's size is an indicator for its primary carriers' tight market PAR. The coefficient for the log of the shipper's annual volume is negative, which implies that larger shippers tend to experience lower tight market PAR. Continuing with model 2a coefficient results, a shipper that is 10 times larger than a fellow shipper experiences tight period PAR that is 1.46\% lower.

In addition, results of models 1a, 2a, and 2b suggest that industry vertical impacts tight market period PAR. Automotive shippers tend to see higher tight market PAR than shippers in the manufacturing industry (recall that the base case industry vertical – i.e., the variable which is omitted from the categorical industry vertical variable in the regression – is Manufacturing). In fact, model 2b indicates that except for those in the paper and packaging industry all industry verticals see higher tight period PAR from non-asst carriers than those in the manufacturing sector.

The numerical illustrations discussed above demonstrate that, while certain tight period shipper behaviors and characteristics impact tight period PAR, in isolation, most of these behaviors may have small impacts on aggregate tight period PAR. In combination, however, they may have greater impact.

In summary, the combination of results that do not support H1a or H1b but do support H2a and H2b, suggest that carriers are myopic: they do not reciprocate good behaviors from shippers in previous soft markets with higher PAR in the following tight market. However they do offer higher PAR in the tight market to shippers that pay higher prices and offer loads more consistently \textit{in the same market period}.

\subsection{Backup carrier price premium}
The preceding results indicate that shippers that pay below market prices in the tight market period observe lower PAR in the same tight periods. These shippers may consider paying more in the tight period to improve PAR. How much the shipper should be willing to pay above its existing contract prices is its Backup Premium: the percent of contract prices the shipper ultimately pays its backup carriers that accept the load (regardless of routing guide depth) or spot carriers as a result of primary carrier rejections:

\begin{equation}
Backup\;Premium = \dfrac{E[Accepted\;Linehaul\;Price | Primary\;Carrier\;Rejects]}{E[Accepted\;Linehaul\;Price | Primary\;Carrier\;Accepts]}
\end{equation}

A Backup Premium value less than 1 indicates contract prices of primary carriers that accept loads are higher than the prices backup carriers charge, or that the shipper often ends up on the spot market and finds capacity at lower prices than contract prices. A Backup Premium of less than 1 may also indicate that loads tend to be rejected by primary carriers on the shippers' lower priced lanes more often than on its higher priced lanes, as the prices used for the Backup Premium are linehaul prices rather than the normalized MRD prices.

The average Backup Premium for shippers in the soft market is 1.009. That is, on average, shippers pay 0.9\% more than their anticipated contract prices as a result of primary carrier rejections in the soft market. However, in the tight market, shippers' average Backup Premium is 1.182. Thus, primary carrier rejections cost shippers 18.2\% more than their contract prices in the tight market. This further demonstrates that not only is primary carrier freight acceptance lower in tight markets, but shippers pay the price for it. This may help shippers negotiate with primary carriers leading into tight markets to increase prices (i.e., MRD), but not by more than their expected Backup Premium.

\subsection{Summary of findings}
Our results suggest that carriers have short memories - they act like goldfish - when market conditions change. Shippers that ``pay it forward'' in soft markets with high pricing relative to the market or consistent volume do not necessarily reap better carrier behavior (i.e. higher PAR) when markets tighten. Put another way, shippers that pay above market prices in the soft market are not more likely to see high PAR in the tight period. In addition, the converse holds: shippers that pay below market prices in soft market periods are not more likely to see PAR decrease as the market tightens. Instead, carriers appear to be myopic. They respond to higher current market period pricing, consistent and frequent load tendering, and low dwell times with higher PAR.

We find that larger shippers are more likely to see lower tight market PAR. This may be explained by the fact that larger shippers interact with more primary carriers, some of which may have a good relationship with the shipper and continue to provide service during tight markets with higher PAR, while other carriers - perhaps those with lower prices or those that serve more volatile lanes or facilities with slower drop off times - may not prioritize the shipper's freight in the same way. Thus, this spread in carrier response to their shippers may be wider for larger shippers and contribute to the negative relationship between shipper size and tight market PAR. While of course this is not a characteristic shippers can change, it is an important confounding factor to include in the model and allow us to understand how different segments of the industry experience market cycles.

\section{Implications}
In this study, we implement a hybrid approach to detect market cycles in TL transportation time series data, which combines structural stability tests and a forward- and backward-looking scoring process. We partition the TL transportation market into two distinct time periods in which the power differential between players shifts from shippers to carriers. Despite prevalent discussions in the literature of the impacts of market conditions on transportation partners, none have considered the impacts of behaviors in one market on behaviors in future market periods.

Our results extend previous literature on S-C relationships, in particular \cite{Scott2017} and \cite{Zsidisin2007} in two ways. First, we verify previous authors' results (each of which consider a single shipper) with our empirical study, which generalizes these findings over a large number of US shippers and carriers across different industry verticals and shipper sizes. Second, we follow individual S-C pairs' interactions over time and include shippers behaviors during markets that favor them in the model of carriers' acceptance decisions when the market is in their favor. We find that there is a propensity for carriers to forget shippers' previous market period behaviors; carriers tend to respond to shippers' behaviors and performance in the current market period rather than based on previously demonstrated behavior.

These results imply that for shippers, it is not necessary to ``pay it forward'' with exceptional performance or pricing in soft markets, but it is important to prioritize particular behaviors in the tight market to ensure high PAR. First, shippers should price competitively relative to market rates. While there are likely diminishing returns to establishing contract prices that are well above market prices, as is suggested both by our empirical analysis and by our industry partner's experience, paying slightly above market prices does result in higher PAR. Shippers may refer to their historical Backup Premiums, particularly at the lane level, to guide the extent of appropriate price increases.

Second, shippers should seek ways to enable consistent tendering behavior. This may include improving forecasting models, contracting a larger set of carriers on particularly volatile lanes (e.g., split the demand on lanes to multiple primary carriers) to smooth out volatility, or bundling lanes for primary carriers such that the volatility of the package of lanes served by each carrier is low. Similarly, shippers can package low volume lanes (i.e., those with low offer cadence) with higher volume lanes to a single carrier so that the cadence of tenders of a set of lanes is more attractive to that carrier.

Finally, shippers' destination facilities must improve dwell times in order to ensure high tight period PAR. The importance of this behavior in particular is underscored by its significance in both the models that test soft market behaviors and tight market behaviors. To reduce dwell times, shippers may improve appointment scheduling, ensure adequate staffing during peak delivery times, expand drop trailer opportunities (as opposed to live unloads, which naturally take longer), or clarify instructions or maps at larger facilities to help drivers find the right loading dock quickly. Driver experiences can make or break shipper reputations across the carrier community. Addressing this problem may be particularly difficult for instances in which loads are being shipped to a downstream customer; the shipper does not own the destination facility and is unable to adjust operations. In these cases, shippers may need to work with their consignees, or recipients, to reduce destination dwell times.

\section{Limitations and Future Research}
As dynamic market conditions impact overall business operations for shippers and carriers, and both sides report interest in developing relationships with one another to ensure better performance, this and future related research have both academic and practical implications. As noted earlier, our model allows us to determine which market-specific behaviors, S-C relationship factors, and shipper and carrier characteristics are important in determining tight market PAR. However, the predictive power is low for our model and as such, we are limited in our ability to quantify direct impact of changing behaviors. Future research may build on these findings and propose a more game theoretic model of shipper-carrier reciprocity to analyze and quantify the impact of specific decisions and behaviors.

A second limitation of this work is that we have limited knowledge of carriers' behaviors outside of the load accept or reject decision. While we know whether the carrier is primary or not with a given shipper, the contracted price, and actual volume tendered and accepted, we do not know what other lanes and shippers the carrier serves exogenous to the given dataset, how well the business realized in our dataset fits into the carriers' overall network, or even expected demand from a shipper on each lane. Shippers' bid cycles occur at effectively random times (from the carriers' perspective) throughout the year and our data includes neither bid timing nor contracted volume. We inherently assume that volume offered to a carrier is approximately that which has been agreed upon during the procurement process. As such, we do not account for surge or unexpected volume from a shipper for which carriers have not planned outside of the effects that are captured by our volatility of volume offered metric. In such cases, our measure of carrier PAR reflects poorly on the carrier as opposed to on the shipper. Future studies may consider how acceptance ratios and routing guide compliance change relative to contracted and unplanned freight volumes.

Finally, in order to retain S-C pairs that interacted in both market periods we removed S-C pairs that interacted in the soft market period but not in the tight market period. There are a multitude of reasons why a S-C pair may interact in one year but not in the future. The shipper may have run a procurement event between the two time periods and for any number of relationship reasons or business fit reasons, the carrier may not have submitted a bid to serve that shipper anymore. This is of course a carrier decision. Or in the same scenario, perhaps the carrier did submit a bid but the shipper chose another carrier to serve the lanes that the first carrier had been serving in the past. This is the shipper’s decision. Finally, if the carrier does not show up in the data again with that shipper, it could be because a load was not tendered to the carrier by the shipper. In this case again, it is the shipper’s decision not to interact with the carrier. However, we don’t have enough information to discern which scenario is or make appropriate assumptions regarding which partner chose to terminate the relationship. As such, we must eliminate these S-C pairs.

The authors of this study will consider a full second soft market period and address the converse research question: do shippers act like elephants or goldfish? Should carriers ``pay it forward'' by adhering to contract commitments in tight periods to ensure shippers reciprocate by upholding their end of the bargain when market soften again?

\renewcommand\refname{References}

\bibliographystyle{plainnat} 
\bibliography{main} 

\pagebreak

\section*{Appendix A: Sensitivity of Jumping Mean method}
\begin{small}
\begin{table}[h]
    \centering
    {Table A.1: Jumping Mean Method break dates by window length}\\
    \smallbreak
    \begin{tabular}{c|c|c|c|c|c}\hline
        6 & 10 & 12 & 26 & 34 & 52 \\\hline\hline
        01-24-16 & 01-24-16 & 02-07-16 & & &\\\hline
        05-29-16 & 05-22-16 & 05-29-16 & & &\\\hline
        08-21-16 & & & 07-24-16 & 06-12-16 &\\\hline
        12-11-16 & 11-13-16 & & &  &\\\hline
        01-22-17 & 01-22-17 & 01-29-17 &  & &\\\hline
        06-04-17 & 06-04-17 & 06-25-17 & 07-23-16 & &\\\hline
        08-27-17 & & & & 09-03-17 & 08-20-17\\\hline
        12-24-17 & 12-03-17 & & & &\\\hline
        01-28-18 & & & & &\\\hline
        03-04-18 & 03-04-18 & & & &\\\hline
        04-29-18 & & & & &\\\hline
        06-10-18 & 06-10-18 & & & &\\\hline
        08-05-18 & & 09-02-18 & & &\\\hline
        10-14-18 & & & & &\\\hline
        01-20-19 & 01-20-19 & 01-20-19 & & &\\\hline
    \end{tabular}
\end{table}
\end{small}
\pagebreak

\section*{Appendix B: Regional price premiums}
Fixed \& per-mile costs: $\hat{\beta}^{Soft}_{base} = \$332.51$; $\hat{\beta}^{Tight}_{base} = \$4.72$; $\hat{\beta}^{Soft}_{dist} = \$1.28/mi$; $\hat{\beta}^{Tight}_{dist} = \$1.36/mi$
All lane distances are over 250 miles (as noted in text), with an average length of haul of 716.4 mi.

\begin{figure*}[h!]
\begin{center}
  \captionsetup{labelformat=empty}
  \caption{Figure B.1: Origin Region Price Premiums, Soft Period}
  \includegraphics[width=0.9\textwidth]{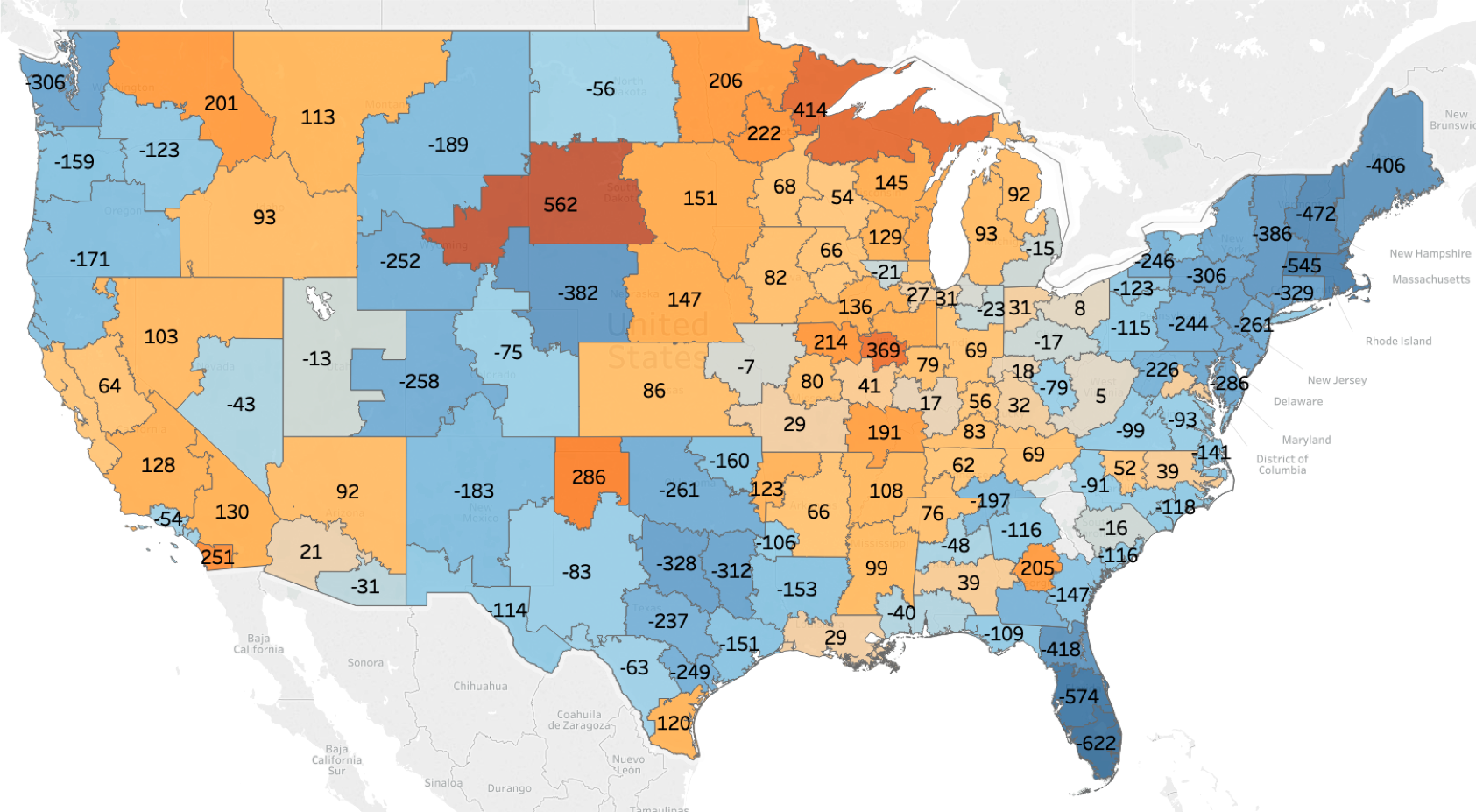}
\end{center}
\end{figure*}

\begin{figure*}[h!]
\begin{center}
  \captionsetup{labelformat=empty}
  \caption{Figure B.2: Origin Region Price Premiums, Tight Period}
  \includegraphics[width=0.9\textwidth]{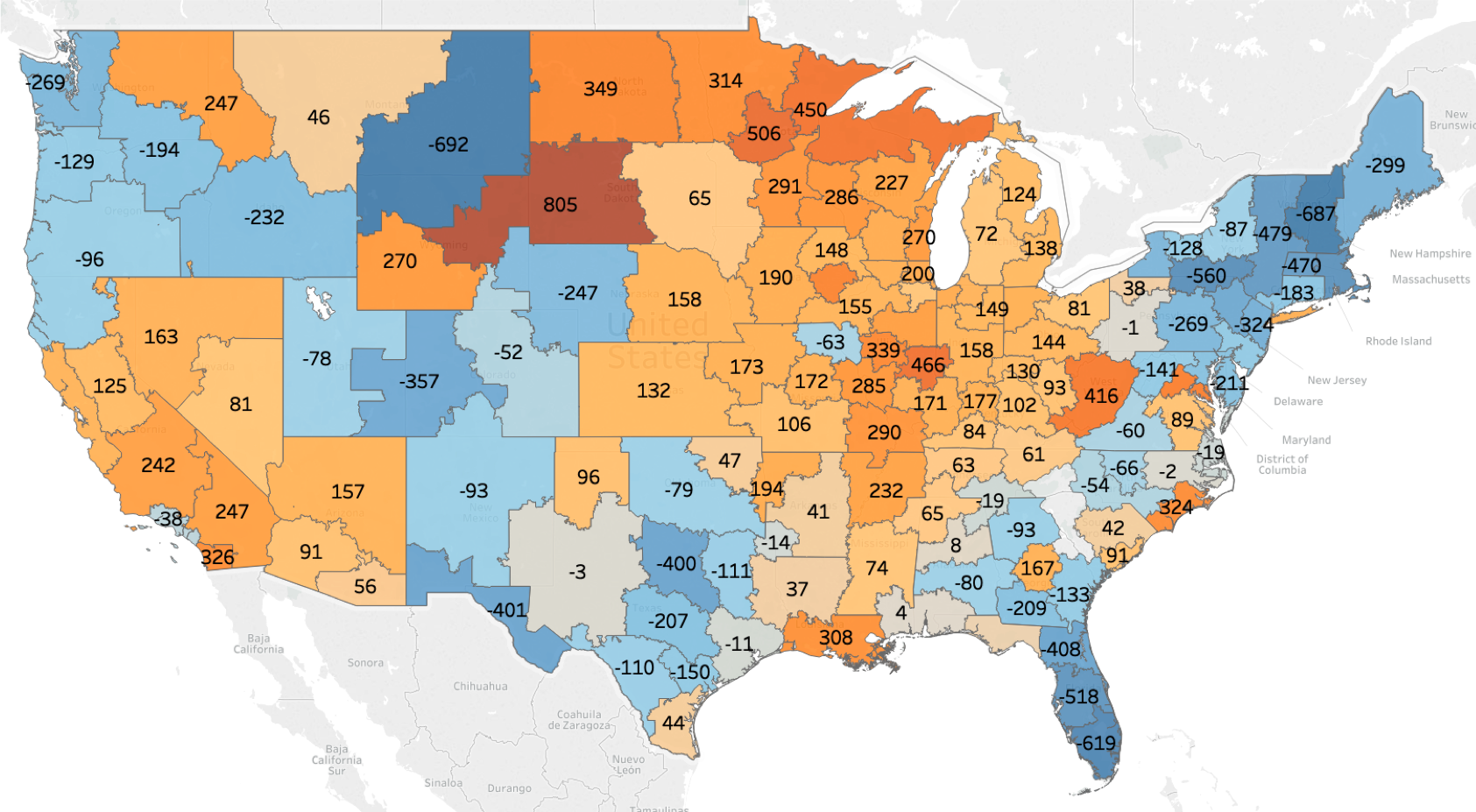}
\end{center}
\end{figure*}

\begin{figure*}[h!]
\begin{center}
  \captionsetup{labelformat=empty}
  \caption{Figure B.3: Destination Region Price Premiums, Soft Period}
  \includegraphics[width=0.9\textwidth]{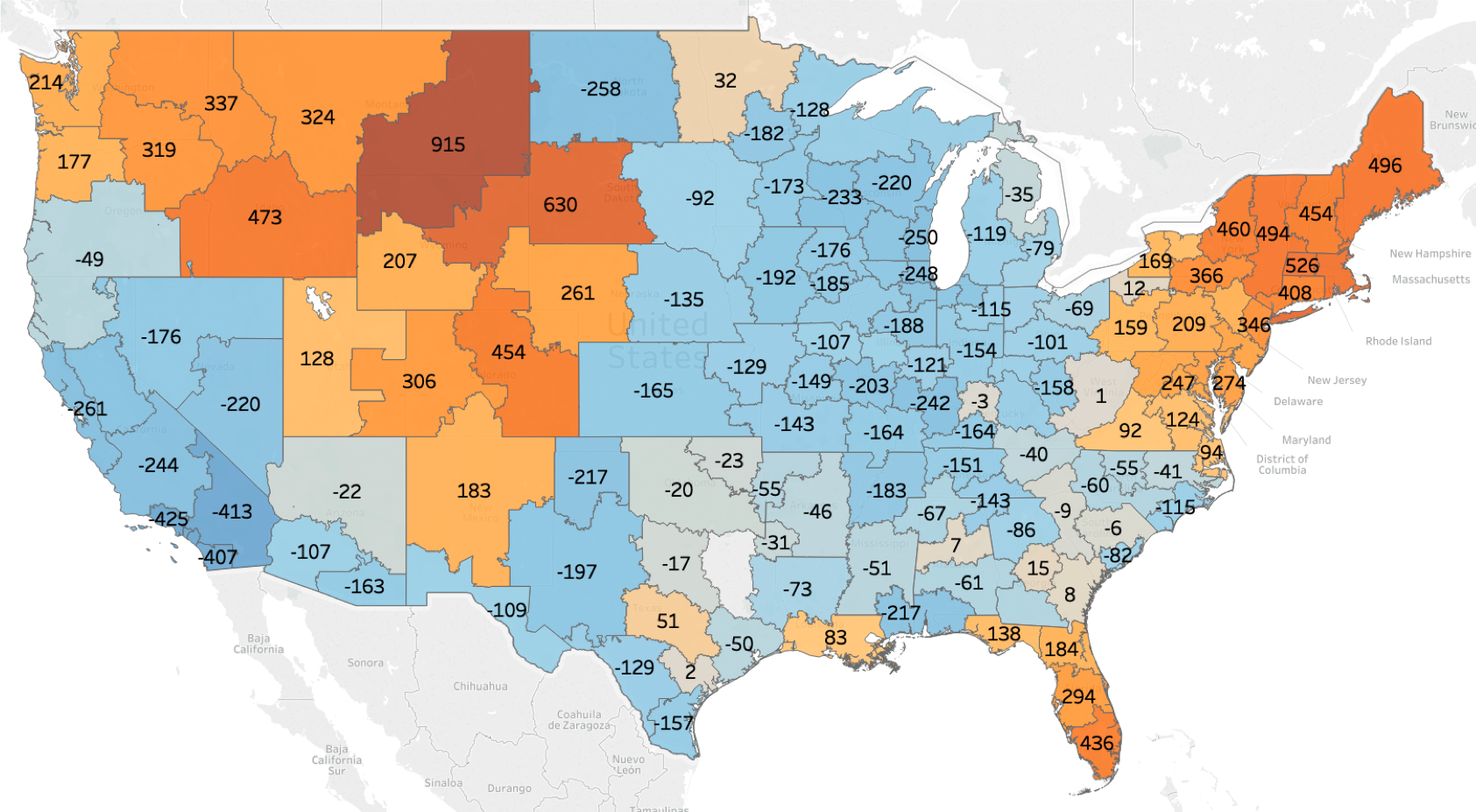}
\end{center}
\end{figure*}
\pagebreak

\begin{figure*}[h!]
\begin{center}
  \captionsetup{labelformat=empty}
  \caption{Figure B.4: Destination Region Price Premiums, Tight Period}
  \includegraphics[width=0.9\textwidth]{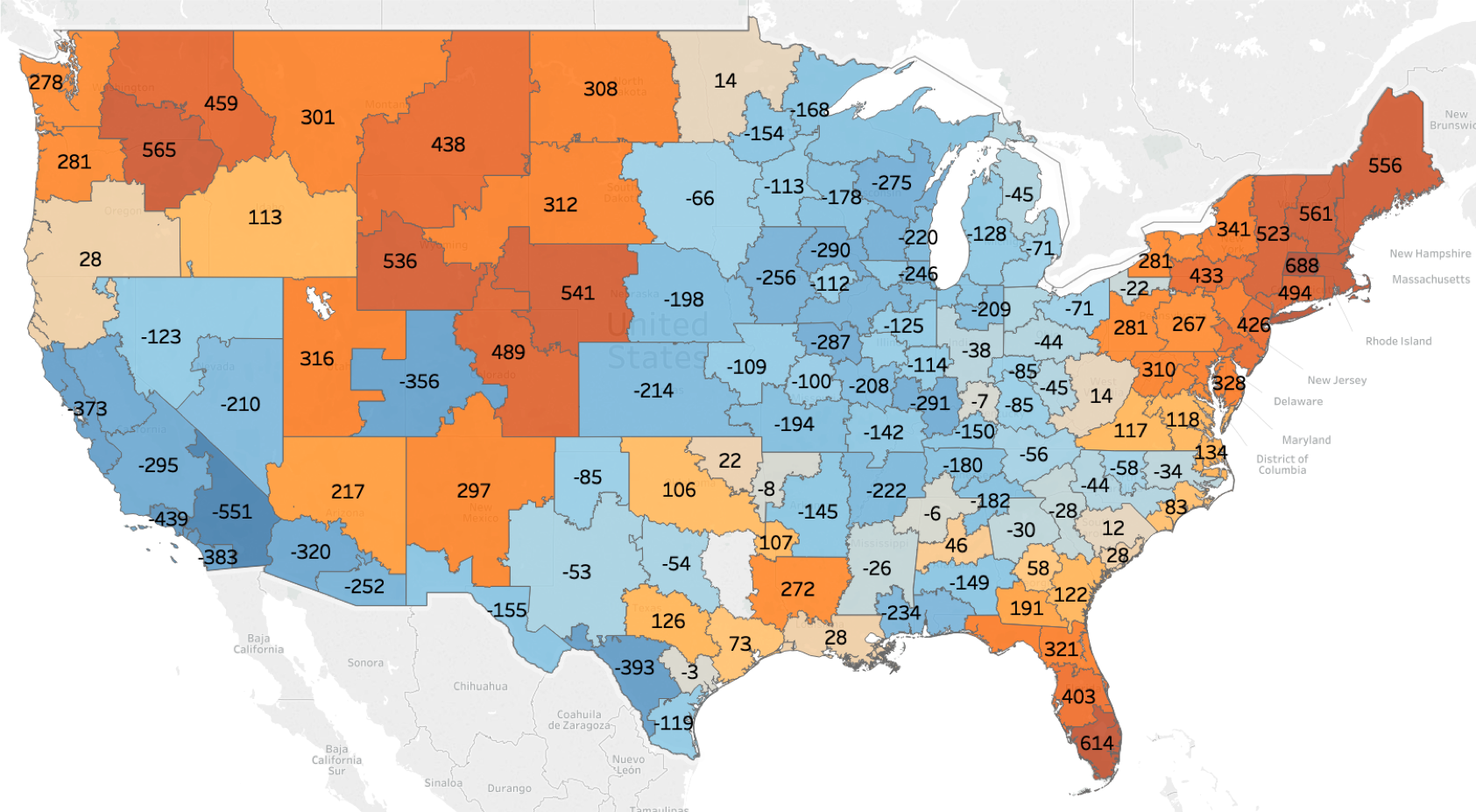}
\end{center}
\end{figure*}

\pagebreak

\section*{Appendix C: Variable summary statistics}
\begin{small}
\begin{centering}
\begin{tabularx}{\linewidth}{l|X||c|c|c|c|c}
 & & & \multicolumn{2}{c|}{Mean} & \multicolumn{2}{c}{St. Dev.} \\
Variable & Definition & N & Soft & Tight & Soft & Tight\\\hline\hline
\multicolumn{2}{l}{Asset-based carriers} & \multicolumn{5}{c}{ }\\\hline
PAR & Average weekly primary carrier acceptance ratio, as a percentage & 503 & 85.9 & 69.0 & 20.6 & 32.2\\\hline
MRD & Market rate differential: the primary carrier's contract rate percentage above or below lane benchmark prices & 503 & 3.06 & -6.12 & 20.91 & 22.25\\\hline
Tendering volatility & Coefficient of variation of weekly tendered volume & 503 & 0.429 & 0.422 & 0.258 & 0.261\\\hline
TLT & Tender lead time (in days) between when load is tendered to primary carrier and when it needs to be picked up & 503 & 4.47 & 4.61 & 2.87 & 3.03\\\hline
Origin dwell & Amount of time (in hours) between when driver arrives at origin empty to leaving with full load & 503 & 2.14 & 2.22  & 1.88 & 2.28\\\hline
Destination dwell & Amount of time (in hours) between when driver arrives at destination with full load to leaving empty or with no trailer & 503 & 1.96 & 2.05 & 1.09 & 1.27\\\hline
Offer cadence & Number of weeks (as a percentage of a 52-week year) in which loads are tendered to the primary carrier & 503 & \multicolumn{2}{c|}{50.40} & \multicolumn{2}{c}{33.21}\\\hline
Shipper size & Log of shipper's total annual volume & 503 & \multicolumn{2}{c|}{3.56} & \multicolumn{2}{c}{0.649}\\\hline
Automotive & Indicator of shipper industry vertical & 159 & \multicolumn{2}{c|}{-} & \multicolumn{2}{c}{-}\\\hline
F\&B/CPG & Indicator of shipper industry vertical & 159 & \multicolumn{2}{c|}{-} & \multicolumn{2}{c}{-}\\\hline
Manufacturing & Indicator of shipper industry vertical & 76 & \multicolumn{2}{c|}{-} & \multicolumn{2}{c}{-}\\\hline
Paper \& Packaging & Indicator of shipper industry vertical & 89 & \multicolumn{2}{c|}{-} & \multicolumn{2}{c}{-}\\\hline
Other & Indicator of shipper industry vertical & 20 & \multicolumn{2}{c|}{-} & \multicolumn{2}{c}{-}\\\hline
Carrier fleet size & Log of the number of trucks (tractors) & 503 & \multicolumn{2}{c|}{2.52} & \multicolumn{2}{c}{0.975}\\\hline\hline
\multicolumn{2}{l}{Non-asset providers} & \multicolumn{5}{c}{ }\\\hline
PAR & (above) & 159 & 88.3 & 69.9 & 18.5 & 32.7\\\hline
MRD & (above) & 159 & 1.49 & -3.52 & 19.15 & 23.7\\\hline
Tendering volatility & (above) & 159 & 0.448 & 0.423 & 0.268 & 0.282\\\hline
TLT & (above) & 159 & 4.16 & 4.24 & 2.72 & 2.51\\\hline
Origin dwell & (above) & 159 & 1.77 & 1.86 & 0.916 & 1.06\\\hline
Destination dwell & (above) & 159 & 1.70 & 0.846 & 1.82 & 1.03\\\hline
Offer cadence & (above) & 159 & \multicolumn{2}{c|}{41.73} & \multicolumn{2}{c}{20.84}\\\hline
Shipper size & (above) & 159 & \multicolumn{2}{c|}{8.26} & \multicolumn{2}{c}{1.46}\\\hline
Automotive & (above) & 41 & \multicolumn{2}{c|}{-} & \multicolumn{2}{c}{-}\\\hline
F\&B/CPG & (above) & 53 & \multicolumn{2}{c|}{-} & \multicolumn{2}{c}{-}\\\hline
Manufacturing & (above) & 32 & \multicolumn{2}{c|}{-} & \multicolumn{2}{c}{-}\\\hline
Paper \& Packaging & (above) & 28 & \multicolumn{2}{c|}{-} & \multicolumn{2}{c}{-}\\\hline
Other & (above) & 5 & \multicolumn{2}{c|}{-} & \multicolumn{2}{c}{-}\\\hline
\end{tabularx}
\end{centering}
\end{small}

\pagebreak
\end{document}